\documentclass[journal]{IEEEtran}

\usepackage[switch]{lineno}

\usepackage{setspace}
\usepackage{balance}
\usepackage{caption}
\usepackage{graphicx}

\usepackage{xcolor,colortbl}

\usepackage{setspace}

\begin{document}
%\linenumbers
\title{Towards Efficient In-memory Computing Hardware for Quantized Neural
Networks: State-of-the-art, Open Challenges and Perspectives}

\author{Olga Krestinskaya, %~\IEEEmembership{Graduate Student Member,~IEEE,}
        Li Zhang, %~\IEEEmembership{Fellow,~OSA,}
        Khaled Nabil Salama %~\IEEEmembership{Senior~Member,~IEEE}% <-this % stops a space
\thanks{
Olga Krestinskaya, Li Zhang, and Khaled Nabil Salama are with the Electrical and Computer Engineering Program, the Division of Computer, Electrical and Mathematical Sciences and Engineering, King Abdullah University of Science and Technology, Thuwal 23955, Saudi Arabia (e-mail: ok@ieee.org; li.zhang@kaust.edu.sa; khaled.salama@kaust.edu.sa).
%GA, 30332 USA e-mail: (see http://www.michaelshell.org/contact.html).
}% <-this % stops a space

}

\maketitle

\begin{abstract}
The amount of data processed in the cloud, the development of Internet-of-Things (IoT) applications, and growing data privacy concerns force the transition from cloud-based to edge-based processing. Limited energy and computational resources on edge push the transition from traditional von Neumann architectures to In-memory Computing (IMC), especially for machine learning and neural network applications. Network compression techniques are applied to implement a neural network on limited hardware resources. Quantization is one of the most efficient network compression techniques allowing to reduce the memory footprint, latency, and energy consumption. This paper provides a comprehensive review of IMC-based Quantized Neural Networks (QNN) and links software-based quantization approaches to IMC hardware implementation. Moreover, open challenges, QNN design requirements, recommendations, and perspectives along with an IMC-based QNN hardware roadmap are provided.

\end{abstract}


% \fi
\IEEEpeerreviewmaketitle

\section{Introduction}

The recently growing data privacy concerns led to the demand to reduce cloud-based processing and to move to local on-edge processing without sharing the data with the server.
Moreover, the development of edge devices and IoT applications created the need to deploy machine learning algorithms and neural networks to low-power devices. 
Also, the neural network models trained on the cloud become larger and more complex. The energy consumption of the data centers to support such AI-related tasks on the cloud is expected to grow exponentially in the next few years \cite{andrae2020new}. 
Therefore, the development and advancement of on-edge processing for both algorithms and hardware are critical.

To move neural network computations to the edge, neural network compression techniques and efficient hardware designs are essential. 
The memory consumed by a state-of-the-art neural network can reach hundreds of megabytes, especially when 32-bit floating-point data representation is used.
%. When a 32-bit floating-point data representation, such networks consume a large amount of memory and can be slow.
Quantization in the neural networks is one of the compression techniques allowing to move from floating point to low-precision fixed-point computations and aiming to reduce the memory footprint, latency, energy consumption, hardware complexity, and computational complexity of a network \cite{gholami2021survey}.

This paper provides a comprehensive review of Quantized Neural Network (QNN) implementations on In-memory Computing (IMC) platforms, related challenges, and open problems. Compared to the previous surveys on QNNs \cite{gholami2021survey, guo2018ASurveyOnMethodsTheoriesofQNN}, this work links software-based QNN designs to IMC hardware implementations, and identifies the related issues. We analyze state-of-the-art IMC-based
QNN implementations, identify open challenges and requirements, and provide recommendations and perspectives. 

The manuscript is organized as follows. Section II focuses on the IMC background, corresponding IMC devices, and state-of-the-art IMC architectures. Section III provides an overview of quantization methods, and QNN training, links quantization schemes with IMC architecture designs, and discusses QNN mapping to IMC hardware. Section IV reviews state-of-the-art IMC implementations of QNN, and compares different quantized IMC designs. Section V discusses the open challenges and requirements for efficient IMC-based QNN hardware along with recommendations and perspectives. Section VI summarises the paper.

\section{In-memory Computing Background}
\IEEEpubidadjcol
The issues of a memory wall and power efficiency ceiling of traditional von Neumann architectures pushed the development of new hardware designs for specific applications, like neural networks. 
Moreover, a large amount of data moved between the memory and processor in von Neumann architectures leads to the demand to search for more efficient alternatives for these applications.
The bottleneck of von Neumann architectures is the bus bandwidth between memory and processor. According to \cite{wang201928}, the bandwidth of this bus reaches 167 GB/s, while the reading operation bandwidth in traditional SRAM memories is 328 TB/s. Also, the data transmission energy between the memory and processor (42 pJ) is 26 times higher than the energy required for the read-out operation (1.6 pJ) \cite{wang201928}.
In-memory Computing (IMC) architecture is an efficient solution to implement Matrix-Vector Multiplication (MVM) operations for machine learning algorithms and neural networks.

\begin{figure*}[!h]
    \centering
    \includegraphics[width=180mm]{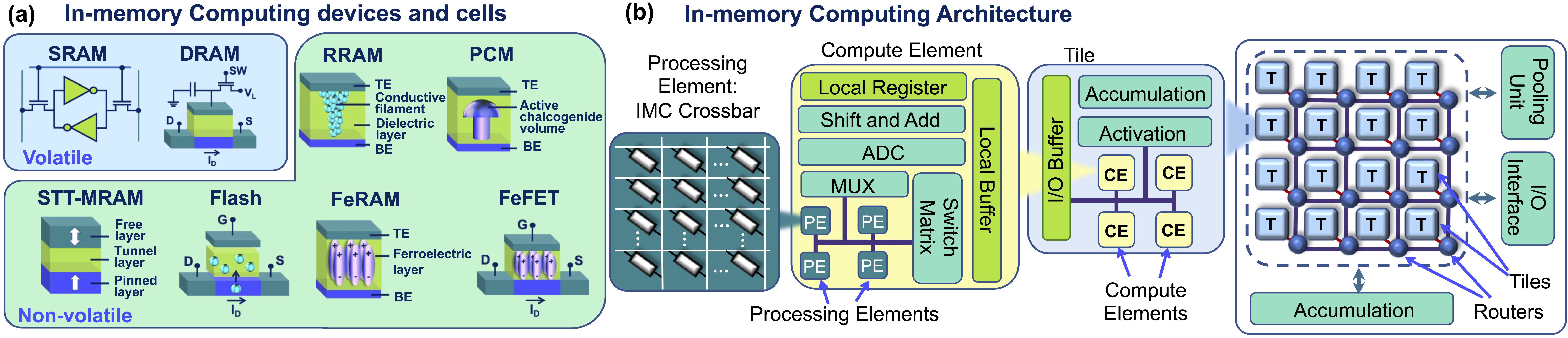}
    \caption{(a) IMC devices \cite{sebastian2020memory, ielmini2020device}. (b) Typical IMC architecture hierarchy \cite{krishnan2021siam}.}
    \label{f1}
\end{figure*}

The devices used for an IMC architecture shown in Fig. \ref{f1} can be divided into volatile, e.g. static random access memory (SRAM) and dynamic random access memory (DRAM), and non-volatile, e.g. resistive random access memories (RRAM), phase change memories (PCM), spin-transfer torque magnetoresistive random access memories (STT-MRAM), ferroelectric RAM (FeRAM), and ferroelectric field-effect transistors (FeFETs) \cite{sebastian2020memory}. SRAM and DRAM memories are more mature, compared to non-volatile memories. Therefore, non-volatile memories, especially RRAMs and PCMs,  suffer from various device non-idealities. These non-idealities include endurance issues, conductance drift, problems of stacking at fault values, and non-linearity of the switching curve. However, non-volatile memories allow multi-level storage, scalability, and high computational density \cite{krestinskaya2022towards}. For example, if the cell size of SRAM is approximately $124F^2$, where $F$ is a technology feature size, RRAM cell with the selector device (1T1R cell) has a size of $12F^2$ \cite{chakraborty2020resistive}.

The IMC devices represent neural network weights and are connected to the crossbar structure to implement IMC architecture, as shown in Fig. \ref{f1} (b). The computation in a crossbar can be performed in the either digital or analog domain. In digital domain computation, the crossbar with IMC devices is used similarly to memory, and the multiplication of each crossbar input with each weight is performed separately. In analog domain computation, the MVM operation is performed in a single cycle, and the voltage inputs to the crossbars are multiplied by the conductance of IMC devices, and the output current through a crossbar column is equivalent to a single entry of MVM.
To perform MVM in the analog domain, the input to the crossbars is applied either through digital-to-analog converters (DACs) for analog inputs, or simple 1-bit comparators when the high-bit inputs are represented by time-encoded binary signals followed by partial sums calculation.
{\color{black}The main problems of a crossbar architecture include sneak path currents \cite{zidan2013memristor}, IR drop \cite{fouda2020ir}, device imperfections \cite{sun2021pcm, krestinskaya2019memristive}, and non-linearity of selector devices usually connected in series with an IMC device \cite{zhang2014selector}.} The typical size of an IMC crossbar is 128x128 or 256x256 \cite{chakraborty2020resistive}. There have been attempts to fabricate larger arrays, however, it increases the effects of sneak path currents and IR drop, especially for non-volatile devices.

The typical IMC architecture is hierarchical (Fig. \ref{f1} (b)). Several crossbar arrays with peripheral circuits, including multiplexers, analog-to-digital converters (ADCs), Shift-and-Add operators, local registers, and control circuits, are connected to form a compute element \cite{krishnan2021siam}.  Several compute elements with the other peripheral circuits, e.g. activation units and buffers, form a tile. In turn, several tiles are connected through network-on-chip (NoC) including routers to direct the signals.

\section{Quantization in IMC-based Neural Network Architectures}
%The main aim of the quantization is to design an efficient neural network architecture. 
%First neural network quantization \cite{fiesler1990FirstQuant}

%Half-precision floating point training \cite{courbariaux2014training}

\subsection{Quantization Methods}
\begin{figure}[!t]
    \centering
    \includegraphics[width=89mm]{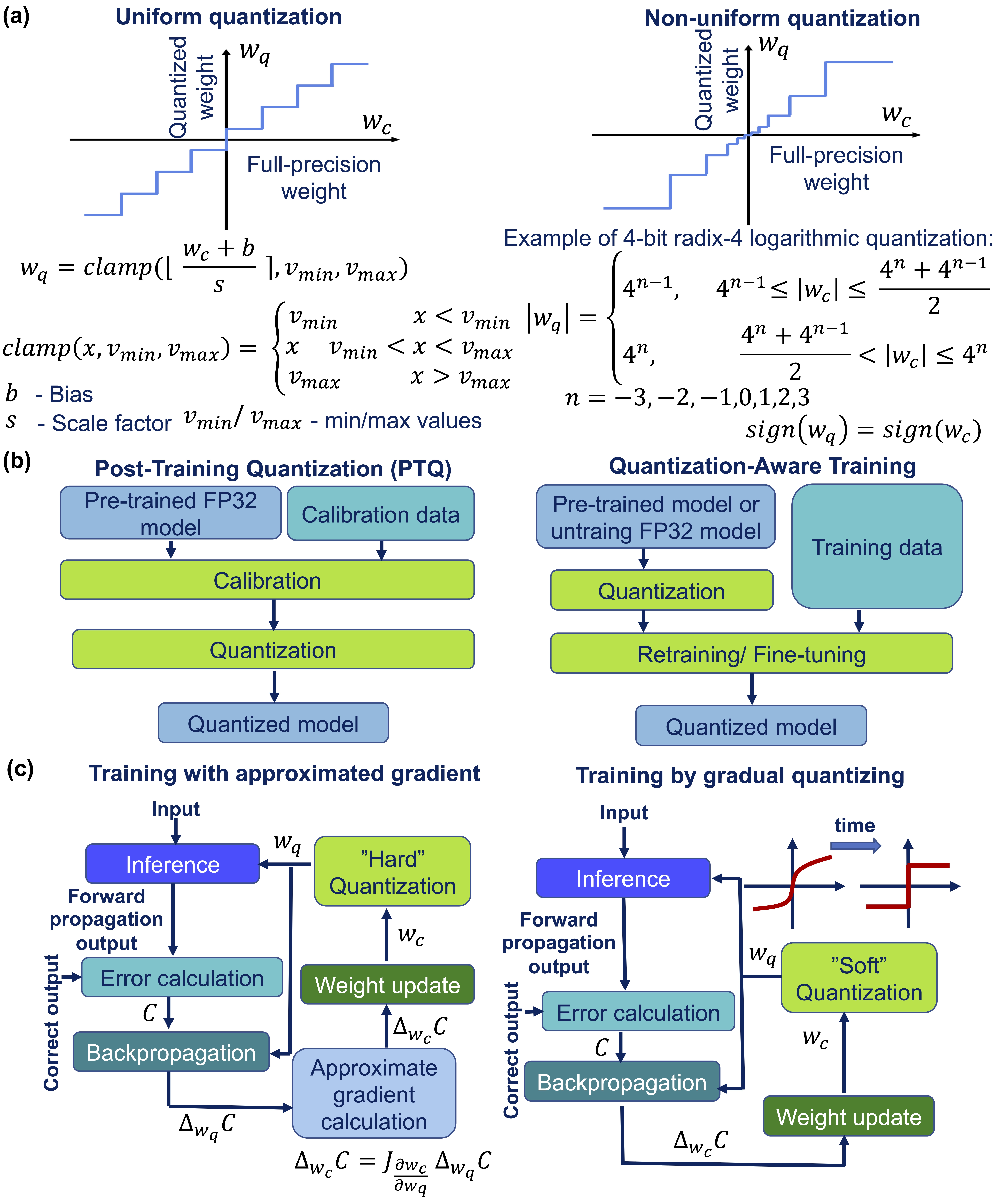}
    \caption{(a) Uniform and non-uniform quantization methods. (b) PTQ vs QAT \cite{gholami2021survey}. (c) QNN training using approximated gradient and gradual quantization.}
    \label{f2}
\end{figure}

Quantizing the neural network implies the quantization of neural network weights, activations, or both.
Quantization methods can be divided into uniform and non-uniform, e.g. logarithmic and codebook quantization.
Uniform quantization is a simple widely-used method, which divides the quantized interval into equally distributed sub-intervals. 
{\color{black} Fig. \ref{f2} (a) illustrates corresponding quantization equations for both methods, where $w_q$ and $w_c$ represent quantized and full-precision weights respectively, $\lfloor x \rceil$ rounds $x$ to the nearest integer and $clamp(x,v_{min},v_{max})$ restricts $x$ within the range of $[v_{min},v_{max}]$.
}
In uniform quantization, the dynamic range of the quantized values is smaller than for non-uniform quantization. To improve this, a layer-wise or channel-wise scaling factor is used at a cost of computational complexity. Non-uniform quantization methods, e.g. logarithmic quantization, have a wider dynamic range {\color{black}allowing better representation of the weights.
In Fig. \ref{f2} (a), 4-bit radix-4 logarithmic quantization is taken as an example of non-uniform quantization, which approximates the absolute value of the weights using $4^n$, $n=-3,-2,-1,...3$ with boundaries being the mid-value of the adjacent values $((4^n+4^{(n-1)})/2)$. 
} 

The other possible quantization method is a codebook quantization, which is useful when the data or weights distribution does not follow linear or logarithmic distributions.  One such method uses k-mean clustering to find the quantized values for a codebook \cite{Gong2014VectorQuant}. The codebook can also be formed by selecting the most frequent values among the weights in a neural network and be updated during training \cite{teng2019LowComplexPolarCodebookQuan}. Reading the codebook can bring additional computational and hardware overhead to neural network implementation.

An important part of  quantization is to determine the clipping range. For neural network weights, the clipping range is static and determined during the training. While, for the activations, it is different, as the inputs change during the inference. Quantization of the activations can be divided into dynamic and static quantization depending on how the clipping of the function is performed. In dynamic quantization, the clipping range is computed dynamically for each activation leading to higher accuracy but more complex computations. In static quantization, the clipping range is pre-calculated \cite{gholami2021survey}.

Different parts of the network exhibit different levels of abstraction and can be affected by the quantization differently \cite{chu2021MPdecreasingBit}. Therefore, it is useful to set different quantization parameters, scale factors, and bit-width for different parts of the network. This is defined as mixed-precision quantization (MPQ). MPQ methods are divided into layer-level and fine-grained MPQs (channel/weight-level) \cite{kim2020MP_Metric_WeightBitLevelChange, nguyen2020MP_Metric_WeightMagnitude}. MPQ parameters can be determined by certain rules, e.g. input layers are more sensitive to the quantization than the other layers in a neural network. Also, MPQ policies can be optimized using differentiable optimization \cite{habi2020HMQ_EDNAS} or reinforcement learning algorithms \cite{elthakeb2019RL_ReLeQ}. MPQ reduces the model size and computation energy demands while keeping high-performance accuracy. However, hardware support and additional hardware overhead are required, especially when fine-grained MPQs are adopted.

\subsection{Quantizing and Training a Neural Network}

There are two main methods to quantize a neural network: Post-Training Quantization (PTQ) and Quantization-Aware Training (QAT) (Fig. \ref{f2} (b)). In PTQ, the well-trained full-precision model is quantized after the training. In QAT, the neural network models are quantized during the training. PTQ is faster but leads to lower accuracy than QAT. The main problem of QAT is a zero-gradient issue due to the stair-like nature of the quantization functions, therefore, the traditional stochastic gradient descent (SGD) algorithm with backpropagation cannot be directly applied for QNN training. 

The issue of zero gradients in QNN training can be solved either by backpropagating approximated gradients or gradually quantizing the network while training (Fig. \ref{f2} (c)). The gradient approximation method is called straight-through estimator (STE), where a Jacobian matrix is set to a diagonal matrix with all 1-s in diagonal entries. The models trained using STE can achieve the accuracy compatible with full-precision models \cite{courbariaux2015binaryconnect}. In general, STE-based methods use full-precision gradients to update the neural network.

The other method to solve the zero-gradient problems in QNN training is a gradual quantization and application of "soft" quantization methods that converge to "hard" quantization over time. One such method is additive noise annealing (ANA) \cite{spallanzani2019ANA}, which injects noise into the quantized variables for the first few iterations of the training. The other method is adding a regularizer to the SGD update \cite{bai2018proxquant}. In addition, to quantize the network gradually, the network can be trained with weights and activations followed by the training with quantized weights and finishing with activation quantization \cite{yang2019QuantizationNetworksAlibaba}.
However, all these training methods require the hardware support of full-precision computation, which makes it challenging to implement on low-power edge devices, especially for IMC architectures.

\subsection{Quantization in In-memory Computing Architectures}

The QNN weights quantization schemes with high bit-precision in a crossbar-based architecture can be either implemented using high-precision devices allowing multi-level storage or combining several fixed-precision n-bit IMC crossbar cells to represent a single weight. 
IMC devices are already quantized by default and can represent several states. For example, in SRAM, the state of the cell is already quantized to two possible levels. In RRAMs, as the number of possible conductance states is also limited, the devices are quantized to a certain precision.
{\color{black} The quantization of the crossbar outputs and activations is defined by ADC and DAC precision \cite{chakraborty2020resistive} and multilevel reading/writing drivers \cite{yilmaz2017drift, ciprut2016modeling}. Moreover, ADC and DAC precision also affects the resolution of neural network weights, e.g. low-precision ADCs cannot capture all the variations of high-precision weights \cite{chakraborty2020resistive}.}

%On the other hand, ADC and DAC precision defines the precision of crossbar outputs and activations and may also affect the resolution of the neural network weights. 

%defines the precision of the activation and affects the precision of 
%The quantization of the activations is defined by ADC and DAC precision. 

In the case of combining several n-bit crossbar cells to form a single weight, the different memory cells represent all bits of the weights from the least significant bit (LSB) to the most significant bit (MSB). The calculation of partial sums is performed to compute the final result (the sum of all the outputs from the different cells), which cause the hardware overhead due to the required ADC resolution. ADC necessary resolution is calculated as $ceil(log_2((2^{b_{DAC}}-1)\times (2^{b_w}-1)\times i))$, where $b_{DAC}$ is a DAC resolution, $i$ is a number of crossbar rows, and $b_w$ is the number of bits in weights \cite{chakraborty2020resistive}. As ADCs are responsible for up to 90\% of the area and power consumption in the compute element \cite{ankit2019puma}, the quantization methods, number of bits in weights and activations, and circuit and architecture design of an IMC core are critical for the hardware efficiency of QNN implementation. Moreover, the partial sums overhead also depends on input slicing and input resolution (activation precision in QNN).

\begin{figure}[!t]
    \centering
    \includegraphics[width=90mm]{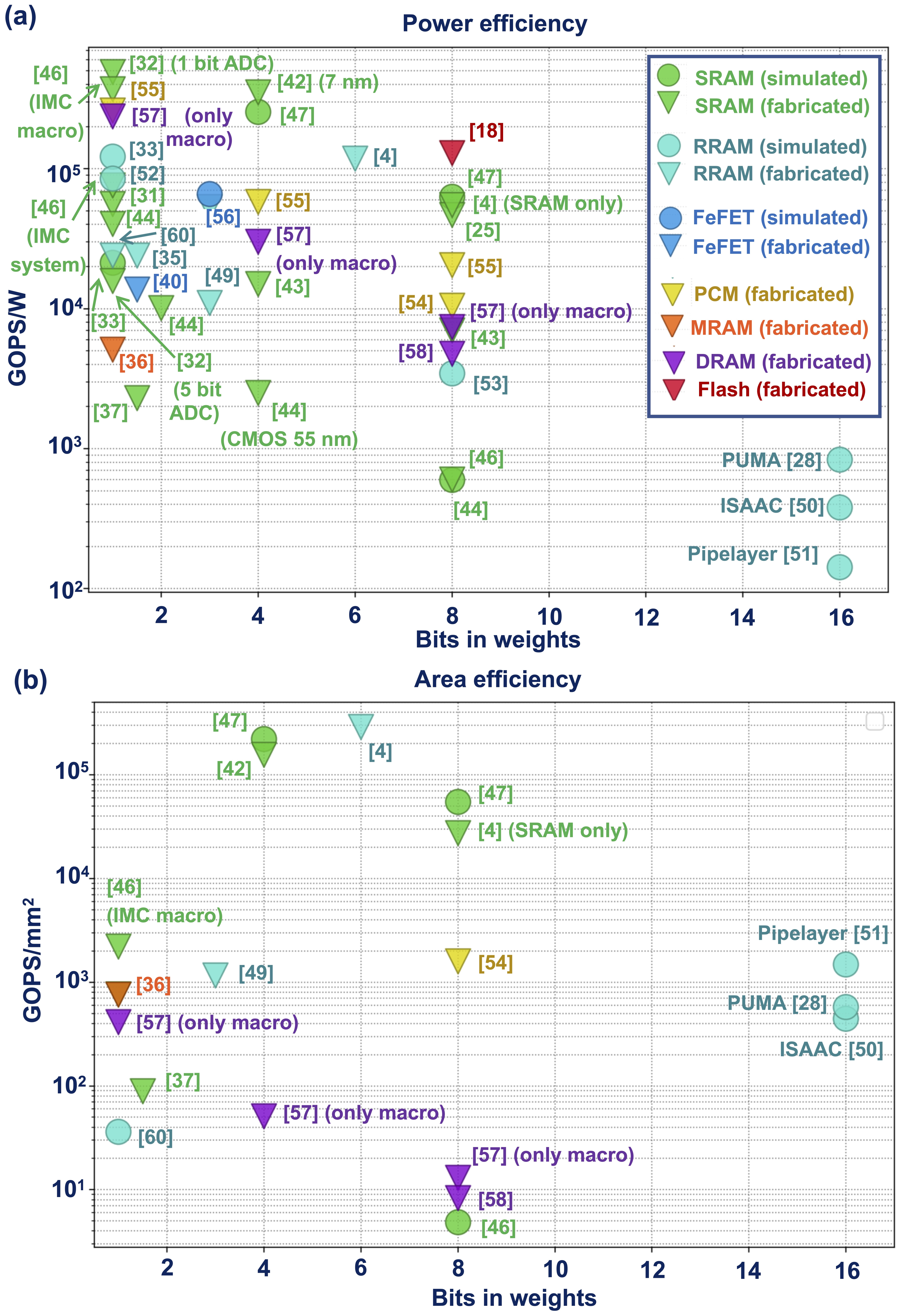}
    \caption{Power and area efficiency of state-of-the-art quantized IMC architectures for AI applications.}
    \label{f3}
\end{figure}

\subsection{Mapping of QNN to IMC architecture}

Mapping of QNN architecture to IMC hardware has several specifications depending on the IMC devices and corresponding hardware designs. 
%If the mapping of SRAM-based implementations have
There are several mapping challenges corresponding to different parts of the design:  non-linear distribution of the quantized levels in IMC devices, quantized weights mapping to several crossbar cells, unrolled convolution kernels mapping to IMC crossbars and mapping of large weight matrices to smaller crossbar arrays.

In high-precision devices, e.g. RRAMs, the quantization levels represented by RRAM conductance are often distributed non-uniformly. Therefore, mapping the uniformly distributed quantized weights to such devices can be challenging, which can also cause a degradation in the performance accuracy.
In quantized high-precision weights mapped to low-precision crossbar cells, the weights can be mapped to several columns in a single crossbar or to several crossbars. This depends on particular designs and should be optimized according to the implemented hardware and availability of peripheral circuits, e.g. ADCs.
In convolution kernels mapping to IMC crossbars to implement CNN, unrolling 3D kernels to vertical columns can cause the circuits overhead, including interconnects and buffers, and dataflow overhead
\cite{peng2019optimizing}. 
In the mapping of large weight matrices to smaller crossbar arrays, the hardware overhead of ADCs should be considered. The search for the optimum crossbar size can be formed as an optimization problem to ensure the efficient mapping of neural network architecture to IMC hardware \cite{negi2021nax}.

\begin{table*}[]
\scriptsize \setlength{\tabcolsep}{12pt}
\renewcommand{\arraystretch}{0.55}
\caption{Summary of QNN hardware for in-memory computing.}
{\color{black}
\begin{tabular}{|lccccl|}
\hline
\multicolumn{1}{|l|}{\textbf{Work}}                                 & \multicolumn{1}{c|}{\textbf{Device}}                                         & \multicolumn{1}{c|}{\textbf{CMOS}} & \multicolumn{1}{c|}{\textbf{Bits (W, A, I)$^*$}} & \multicolumn{1}{c|}{\textbf{\begin{tabular}[c]{@{}c@{}}Implemented\\ network\end{tabular}}} & \multicolumn{1}{c|}{\textbf{Details}}                              \\ \hline
\multicolumn{6}{|c|}{\textbf{Binary designs}}                                                                                                                                                                                                                                                                                                                                                            \\ \hline
\multicolumn{1}{|l|}{\cite{khwa201865nm}}          & \multicolumn{1}{c|}{SRAM}                                                    & \multicolumn{1}{c|}{65nm}          & \multicolumn{1}{c|}{1, 1.5$^{*1}$, 1}             & \multicolumn{1}{c|}{CNN}                                                                    & Fabricated SRAM-based IMC macro                                    \\ \hline
\multicolumn{1}{|l|}{\cite{yu202016k}}             & \multicolumn{1}{c|}{SRAM}                                                    & \multicolumn{1}{c|}{65nm}          & \multicolumn{1}{c|}{1, 1-5, 1}              & \multicolumn{1}{c|}{3-layer ANN$^{*2}$}                                                          & Fabricated SRAM bitcell array for BNN                              \\ \hline
\multicolumn{1}{|l|}{\cite{sun2018computing}}      & \multicolumn{1}{c|}{\begin{tabular}[c]{@{}c@{}}SRAM/\\ RRAM$^{*3}$\end{tabular}} & \multicolumn{1}{c|}{65nm}          & \multicolumn{1}{c|}{1, 1, 1}               & \multicolumn{1}{c|}{VGG}                                                                    & Simulated XNOR-based BNN                                           \\ \hline
\multicolumn{1}{|l|}{\cite{sun2018fully}}          & \multicolumn{1}{c|}{RRAM}                                                    & \multicolumn{1}{c|}{65nm}          & \multicolumn{1}{c|}{1, 1, 1}               & \multicolumn{1}{c|}{5-layer CNN}                                                            & Simulated fully-parallel BNN                                       \\ \hline

\multicolumn{1}{|l|}{\cite{yin2020high}}           & \multicolumn{1}{c|}{RRAM}                                                    & \multicolumn{1}{c|}{90nm}          & \multicolumn{1}{c|}{1, 1, 1}               & \multicolumn{1}{c|}{\begin{tabular}[c]{@{}c@{}}9-layer CNN, \\ 5-layer ANN\end{tabular}}    & Fabricated XNOR-RRAM prototype chip                                \\ \hline
\multicolumn{1}{|l|}{\cite{deaville2021maximally}} & \multicolumn{1}{c|}{MRAM}                                                    & \multicolumn{1}{c|}{22nm}          & \multicolumn{1}{c|}{1, 4, 1}               & \multicolumn{1}{c|}{6-layer CNN}                                                            & Fabrucated macro with high parallelism                             \\ \hline
\multicolumn{6}{|c|}{\textbf{Ternary designs}}                                                                                                                                                                                                                                                                                                                                                           \\ \hline
\multicolumn{1}{|l|}{\cite{ando2017brein}}         & \multicolumn{1}{c|}{SRAM}                                                    & \multicolumn{1}{c|}{65nm}          & \multicolumn{1}{c|}{1.5, 1.5, -}            & \multicolumn{1}{c|}{13-layer DNN$^{*2}$}                                                         & Reconfigurable accelerator with ternarized mask                    \\ \hline
\multicolumn{1}{|l|}{\cite{laborieux2020low}}      & \multicolumn{1}{c|}{RRAM}                                                    & \multicolumn{1}{c|}{130nm}         & \multicolumn{1}{c|}{1.5, 1.5, 1.5}          & \multicolumn{1}{c|}{VGG}                                                                    & Gated XNOR with near-throushold sense amplifier                    \\ \hline
\multicolumn{1}{|l|}{\cite{zhu2022fat}}            & \multicolumn{1}{c|}{MRAM}                                                    & \multicolumn{1}{c|}{45nm}          & \multicolumn{1}{c|}{1.5, 8, -}              & \multicolumn{1}{c|}{CNN}                                                                    & Sparse and fast additions for IMC                                  \\ \hline

\multicolumn{1}{|l|}{\cite{saito2021analog}}       & \multicolumn{1}{c|}{FeFET}                                                   & \multicolumn{1}{c|}{22nm}          & \multicolumn{1}{c|}{1.5, 4, -}              & \multicolumn{1}{c|}{ResNet20}                                                               & Analog crossbar with FeFET and tunnel junction resistor            \\ \hline
\multicolumn{6}{|c|}{\textbf{Higher precision designs}}                                                                                                                                                                                                                                                                                                                                                  \\ \hline
\multicolumn{1}{|l|}{\cite{yang201924}}            & \multicolumn{1}{c|}{SRAM}                                                    & \multicolumn{1}{c|}{28nm}          & \multicolumn{1}{c|}{8,8,8}                  & \multicolumn{1}{c|}{AlexNet}                                                                & CNN with sandwich-shaped SRAM                                      \\ \hline
\multicolumn{1}{|l|}{\cite{wang201928}}            & \multicolumn{1}{c|}{SRAM}                                                    & \multicolumn{1}{c|}{28nm}          & \multicolumn{1}{c|}{1-8$^{*4}$}                & \multicolumn{1}{c|}{CNN}                                                                    & Hybrid in-/near-memory compute SRAM for IMC                        \\ \hline
\multicolumn{1}{|l|}{\cite{dong202015}}            & \multicolumn{1}{c|}{SRAM}                                                    & \multicolumn{1}{c|}{7nm}           & \multicolumn{1}{c|}{4, 4, 4}               & \multicolumn{1}{c|}{-}                                                                      & Fabricated reconfigurable FinFET SRAM IMC macro                    \\ \hline
\multicolumn{1}{|l|}{\cite{su2021two}}             & \multicolumn{1}{c|}{SRAM}                                                    & \multicolumn{1}{c|}{28nm}          & \multicolumn{1}{c|}{4-8, 2-8, 10-20}        & \multicolumn{1}{c|}{ResNet-20}                                                              & SRAM-based inference and training accelerator                      \\ \hline

\multicolumn{1}{|l|}{\cite{zhang201955nm}}         & \multicolumn{1}{c|}{SRAM}                                                    & \multicolumn{1}{c|}{55nm}          & \multicolumn{1}{c|}{1-8}                    & \multicolumn{1}{c|}{-}                                                                      & Configurable hybrid SRAM macro                                     \\ \hline
\multicolumn{1}{|l|}{\cite{yue202115}}             & \multicolumn{1}{c|}{SRAM}                                                    & \multicolumn{1}{c|}{65nm}          & \multicolumn{1}{c|}{1-8, 2-8, -}            & \multicolumn{1}{c|}{VGG, ResNet}                                                            & IMC macro with zero-activation and zero-weight skipping            \\ \hline
\multicolumn{1}{|l|}{\cite{guo202115}}             & \multicolumn{1}{c|}{SRAM}                                                    & \multicolumn{1}{c|}{28nm}          & \multicolumn{1}{c|}{2-4, 4-8, -}            & \multicolumn{1}{c|}{RebNet-20, LSTM}                                                        & IMC macro with variable precision quantization                     \\ \hline
\multicolumn{1}{|l|}{\cite{fujiwara20225}}         & \multicolumn{1}{c|}{SRAM}                                                    & \multicolumn{1}{c|}{5nm}           & \multicolumn{1}{c|}{4, 4, 14}               & \multicolumn{1}{c|}{-}                                                                      & Digital IMC macro                                                  \\ \hline

\multicolumn{1}{|l|}{\cite{wang2019situ}}          & \multicolumn{1}{c|}{RRAM}                                                    & \multicolumn{1}{c|}{-}             & \multicolumn{1}{c|}{8, - , 16}              & \multicolumn{1}{c|}{CNN, ConvLSTM}                                                          & Fabricated network with on-chip learning                           \\ \hline
\multicolumn{1}{|l|}{\cite{yao2020fully}}          & \multicolumn{1}{c|}{RRAM}                                                    & \multicolumn{1}{c|}{130nm}         & \multicolumn{1}{c|}{3, 8 , 8$^{*5}$}             & \multicolumn{1}{c|}{5-layer CNN}                                                            & Fabricated fully implemented network with on-chip fine-tuning      \\ \hline
\multicolumn{1}{|l|}{\cite{shafiee2016isaac}}      & \multicolumn{1}{c|}{RRAM}                                                    & \multicolumn{1}{c|}{32nm}          & \multicolumn{1}{c|}{16, 16, 16$^{*5}$}           & \multicolumn{1}{c|}{VGG}                                                                    & Full pipelined accelerator architecture ISAAC                      \\ \hline
\multicolumn{1}{|l|}{\cite{song2017pipelayer}}     & \multicolumn{1}{c|}{RRAM}                                                    & \multicolumn{1}{c|}{-}             & \multicolumn{1}{c|}{16, 16, 16$^{*5}$}           & \multicolumn{1}{c|}{AlexNet, VGG}                                                           & Full accelerator architecture Pipelayer for training and inference \\ \hline
\multicolumn{1}{|l|}{\cite{ankit2019puma}}         & \multicolumn{1}{c|}{RRAM}                                                    & \multicolumn{1}{c|}{32nm}          & \multicolumn{1}{c|}{16, 16, 16$^{*5}$}           & \multicolumn{1}{c|}{VGG, LSTM}                                                              & Full programmable accelerator architecture PUMA with compiler      \\ \hline

\multicolumn{1}{|l|}{\cite{li2021secure}}          & \multicolumn{1}{c|}{RRAM}                                                    & \multicolumn{1}{c|}{40nm}          & \multicolumn{1}{c|}{1-8, 3, -}              & \multicolumn{1}{c|}{VGG-8}                                                                  & Reconfigurable IMC macro with sparsity control                     \\ \hline
\multicolumn{1}{|l|}{\cite{zhu2019configurable}}   & \multicolumn{1}{c|}{RRAM}                                                    & \multicolumn{1}{c|}{45nm}          & \multicolumn{1}{c|}{-, 6-10, -}             & \multicolumn{1}{c|}{\begin{tabular}[c]{@{}c@{}}Lenet, VGG-16,\\ ResNet-18\end{tabular}}     & IMC macro with configurable precision and layer-wise quantization  \\ \hline
\multicolumn{1}{|l|}{\cite{khaddam2021hermes}}     & \multicolumn{1}{c|}{PCM}                                                     & \multicolumn{1}{c|}{14nm}          & \multicolumn{1}{c|}{8, -, 8}                & \multicolumn{1}{c|}{ANN, ResNet-9}                                                          & IMC macro with local digital processing                            \\ \hline
\multicolumn{1}{|l|}{\cite{khwa202240}}            & \multicolumn{1}{c|}{PCM}                                                     & \multicolumn{1}{c|}{40nm}          & \multicolumn{1}{c|}{2-8, 5-19, 1-8}         & \multicolumn{1}{c|}{ResNet-20}                                                              & IMC with PCM macro                                                 \\ \hline
\multicolumn{1}{|l|}{\cite{matsui2021energy}}      & \multicolumn{1}{c|}{FeFET}                                                   & \multicolumn{1}{c|}{-}             & \multicolumn{1}{c|}{3,-,-}                  & \multicolumn{1}{c|}{-}                                                                      & FeFET-based IMC with charge sharing                                \\ \hline
\multicolumn{1}{|l|}{\cite{xie2022gain}}           & \multicolumn{1}{c|}{DRAM}                                                    & \multicolumn{1}{c|}{65nm}          & \multicolumn{1}{c|}{1-8, 2-16, 2-8}         & \multicolumn{1}{c|}{ResNet-20}                                                              & Gain-cell eDRAM for IMC                                            \\ \hline
\multicolumn{1}{|l|}{\cite{xie202116}}             & \multicolumn{1}{c|}{DRAM}                                                    & \multicolumn{1}{c|}{65nm}          & \multicolumn{1}{c|}{8,8,8}                  & \multicolumn{1}{c|}{6-layer CNN}                                                            & Charge-based eDRAM for IMC                                         \\ \hline
\multicolumn{1}{|l|}{\cite{kim2021embedded}}       & \multicolumn{1}{c|}{Flash}                                                   & \multicolumn{1}{c|}{65nm}          & \multicolumn{1}{c|}{8,-,8}                  & \multicolumn{1}{c|}{LeNet-5}                                                                & 3-D NAND Flash IMC array                                           \\ \hline
\multicolumn{6}{|l|}{$^*$: W,A,I - weights, activations, inputs,  $^{*1}$: ternary, $^{*2}$: fully-connected, $^{*3}$: different architectures,  $^{*4}$: several cases, $^{*5}$: serial 1-bit input. }                                                                                                                                                                                                                                                                                                                                                            \\ \hline
\end{tabular} }
\label{tabnew}
\end{table*}

\section{In-memory Computing Hardware for QNN}

\subsection{Binarized Neural Networks and QNNs with Ternary Weights}

Binarized neural network (BNN) is a type of QNN relying on 1-bit weights (+1 and -1 in software) and activations \cite{krestinskaya2022towards}.
The fabricated binary SRAM-based IMC designs are shown in \cite{khwa201865nm,yu202016k}.
The efficient implementation of BNN in IMC hardware is based on XNOR operation, which reduces ADC overhead using 1-bit sense amplifiers \cite{sun2018fully} for binarized activations. 
XNOR-based implementation can be 30 times more efficient than sequential row-by-row read-out.
Comparing the implementations of binarized SRAM- and RRAM- based networks, RRAM-based implementation is 5.8 times more energy-efficient than 8T SRAM-based design \cite{sun2018computing}.
The other RRAM-based fabricated prototype of BNN with flash ADCs is shown in \cite{yin2020high}. The application of flash ADCs is acceptable for BNNs, however for higher-bit designs, flash ADCs can bring significant area and power overhead \cite{krestinskaya2022towards}.
A fabricated MRAM-based IMC macro for 1-bit operations is shown in \cite{deaville2021maximally}.

The QNN designs with ternary weights imply the quantized values represented by -1, 0, and +1. {\color{black} Ternary weights networks demonstrate higher accuracy than BNNs and better sparsity, which allows skipping the operations related to zero weights \cite{zhu2022fat}. Ternary networks} can be implemented using RRAM-based crossbar arrays, where positive and negative weights are represented by two crossbar cells, and a subtractor is used to calculate the final value of the weight, which allows getting zero-weights \cite{liu202033}. The implementation of a ternarized accelerator using SRAMs is also possible \cite{ando2017brein}. {\color{black} In \cite{ando2017brein}, the network is ternarized introducing a mask. }This design aims for reconfigurability for different types of neural network architectures.
{\color{black} \cite{zhu2022fat} demonstrates STT-MRAM based ternary neural network implementation with sparse and fast addition.}
The ternary weights in FeFET-based architecture are illustrated in \cite{saito2021analog}.
{\color{black} \cite{laborieux2020low} illustrates the ternary XNOR network with RRAM devices.}

\subsection{Higher-Bit Fixed-point IMC Computations for QNNs}

The most common IMC architectures for fixed-point computations are based on SRAM and RRAM memories.
As SRAM is a mature memory technology, there are a lot of fabricated SRAM-based IMC architectures shown in \cite{yang201924, wang201928, dong202015}. The SRAM implementation of both inference and training accelerator is illustrated in \cite{su2021two}. 
The fabricated RRAM-based IMC macros designs are shown in \cite{wang2019situ, yao2020fully}. In \cite{wang2019situ}, on-chip learning is considered. In \cite{yao2020fully}, on-chip fine-tuning is demonstrated.
The RRAM-based IMC accelerator level designs with several levels of architecture hierarchy are shown in \cite{shafiee2016isaac,song2017pipelayer,ankit2019puma,ankit2020panther}. These QNN designs include the datapath, routing, and arrangement of the network nodes are considered in addition to controlling and computation peripherals \cite{krestinskaya2022towards}.

The general trend to move towards general-purpose implementations in QNN accelerators rather than architecture-specific designs leads to the development of configurable bit-precision in IMC architectures.
Several works are implementing configurable macros for IMC accelerators to be able to vary the bit precision of the weights \cite{zhang201955nm,yue202115, guo202115, fujiwara20225}. Some recent RRAM-based IMC architectures can also support configurable bit-operations \cite{li2021secure, zhu2019configurable}.
Apart from SRAM and RRAM, the other IMC devices are less popular for IMC hardware. However, there are several architectures based on PCMs \cite{khaddam2021hermes, khwa202240}, FeFETs \cite{matsui2021energy}, DRAMs \cite{xie2022gain, xie202116}, and flash memories \cite{kim2021embedded}.

%\scriptsize \setlength{\tabcolsep}{12pt}
%\renewcommand{\arraystretch}{0.55}

%PCM-based IMC architectures are shown in .
%FeFET-based macro is demonstrated in 
%The DRAM-based IMC accelerators with variable bit precision are shown in . The disadvantage of DRAM is a destructive read operation.
%NAND Flash-memory-based IMC architecture is illustrated in \cite{kim2021embedded}.

{\color{black}Table \ref{tabnew} shows the summary QNN hardware designs for IMC.} Fig. \ref{f3} illustrates the comparison of power and area efficiency of different IMC hardware designs depending on the weights precision {\color{black} for the architectures in Table \ref{tabnew}. }
It is important to note that ADC precision affects power efficiency significantly. In \cite{yu202016k}, the design with 1-bit ADC is 30 times more efficient than for 5-bit ADC. 
The precision of weights also affects power efficiency. In \cite{zhang201955nm}, the power efficiency drops from 40.2 TOPS/W to 0.6 TOPS/W for 1-bit and 8-bit weights accordingly.
The power efficiency of the IMC macro and the power efficiency of the whole system including several macros connected with peripheral circuits has a significant difference, e.g. in \cite{yue202115} IMC macro is 5 times more efficient than the overall system.
Power efficiency is also affected by the technology node. For example, for a 4-bit neural network, 7nm SRAM-based IMC architecture \cite{dong202015} is more than 100 times more efficient than 55nm SRAM-based architecture \cite{zhang201955nm}.

\begin{figure*}[t]
    \centering
    \includegraphics[width=180mm]{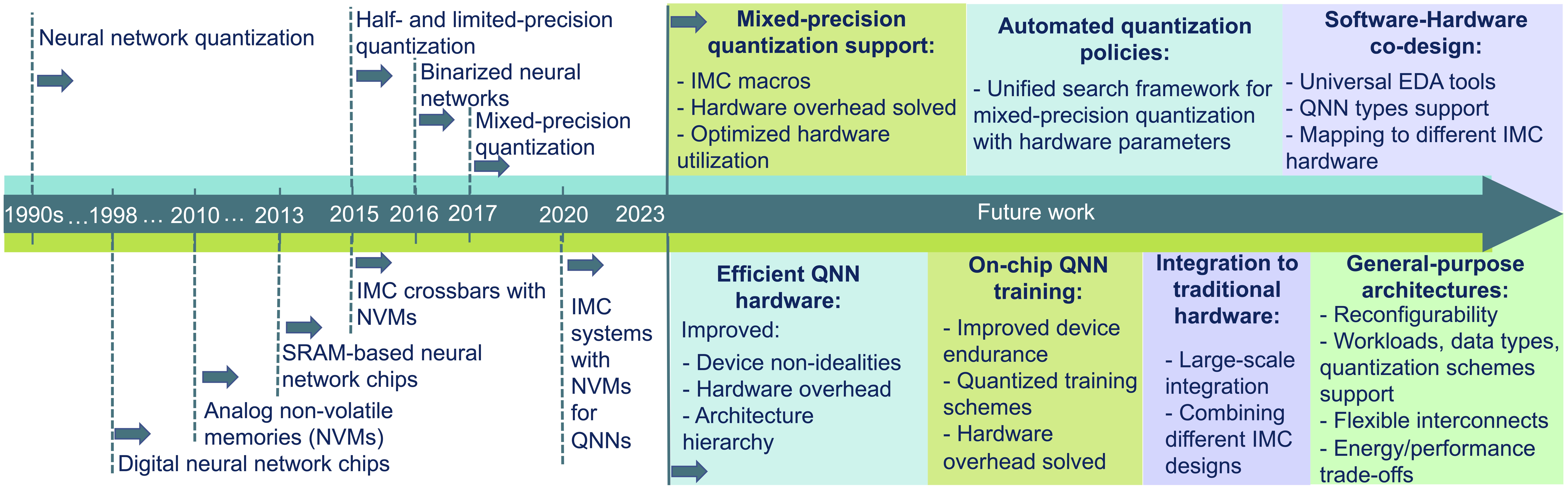}
    \caption{Roadmap for IMC-based QNN hardware (the ideas of software-hardware co-design and general-purpose architectures are adopted from \cite{zhang2020neuro} and extended for QNN implementation).}
    \label{roadmap}
\end{figure*}

\section{Open Challenges, Requirements, Recommendations, and Perspectives}

The roadmap for IMC-based QNN architectures with recent achievements and open challenges is illustrated in Fig. \ref{roadmap}.  The recommendations for future development of IMC-based QNN hardware include general improvement in QNN hardware designs, on-chip training, mixed-precision quantization and design reconfigurability support, automation of the quantization policies search, development of software-hardware co-design frameworks, and integration of IMC architectures to the traditional designs.

\subsection{Efficient QNN Inference Architecture}

Despite the recent development of IMC inference hardware designs, there are still challenges that should be addressed to move from IMC accelerators developed in research laboratories to efficient commercial solutions for low-power IMC-based QNN hardware.
They include the hardware overhead in crossbar macros, consideration of architecture hierarchy and related challenges, and hardware non-idealities affecting performance accuracy.

Even though IMC crossbar architecture can be highly efficient for MVM operations, the hardware overhead caused by peripheral and control circuits can minimize the benefits of even small low-power non-volatile memories in the design. In crossbar macros, the hardware overhead comes from
ADC and partial sums, when high-precision weights split into low-precision crossbar cells \cite{chakraborty2020resistive}.
ADC design improvements, implementation of low-power converters, reducing the number of ADCs and ADC resolution, and relying on approximate computing can benefit IMC hardware designs \cite{krestinskaya2022towards}. 

An efficient IMC hierarchy design is also important. The interconnection of processing elements, computation blocks, and tiles also affects design efficiency. The architecture hierarchy in SRAM-based IMC designs is explored more than in non-volatile memory-based designs. Data movement between the layers, requirements for additional storage, and circuits interconnection in each level of the IMC architecture hierarchy should still be improved further.
Moreover, several solutions have been explored to move from a traditional monolithic chip design to 2.5D integration or chiplet-based designs \cite{krishnan2021siam}.

 IMC designs based on non-volatile devices are prone to non-idealities. Even though quantized and binarized architectures are affected less by noise and device variations \cite{james2022inference}, device-to-device and cycle-to-cycle variability cause errors propagating through the network and affecting the performance accuracy. In addition, the immaturity of non-volatile memories leads to device fault issues and conductance drift with time also reducing performance accuracy \cite{krestinskaya2022towards}. These issues along with the device endurance should still be addressed at the device and material level. 
Moreover, the 3D stacking capability of IMC devices and crossbar cells should also be developed, as
3D integration is useful to decrease the length of interconnect wires to increase the chip density and reduce IR drop.

\subsection{QNN Training on Chip}

One of the main problems of QNN training is the requirement of full-precision gradients for weight updates. This leads to the complexity of neural network training on low-power hardware and IMC architectures that do not support full-precision operations. Several techniques to reduce the number of full-precision weights, e.g. mixing the full-precision weights with quantized weights for training \cite{chen2020nBitQNN}. However, the hardware support for full-precision operation is still necessary in this case. Therefore, the algorithms for QNN training based only on quantized values still need to be developed.  
Moreover, it is important to adapt these algorithms for IMC hardware relying on hardware-friendly operations, and include IMC hardware considerations, e.g. ADC precision considerations, partial sums,  and  IMC hardware non-idealities.

Neural network training on a chip can also be limited by the device non-idealities. 
If IMC hardware non-idealities are not the main concern in SRAM-based architectures, RRAM-based IMC architectures suffer from endurance issues and variabilities.  The endurance of  $10^8-10^9$ cycles is required to implement on-chip training, while most of the current RRAM devices are limited to $10^6$ cycles
\cite{chakraborty2020resistive}. The other issue related to on-chip training implementation of non-volatile IMC devices, e.g. RRAMs and PCMs, is a high write latency. The write latency of RRAMs and PCMs usually reaches 100-150ns, and the write latency of SRAMs can be less than 1ns \cite{haensch2022co}. Non-linearity in the switching behavior of some non-volatile IMC devices, such as RRAMs, also requires additional hardware overhead to verify and correct the programmed conductance values for on-chip training \cite{krestinskaya2022towards}.

On-chip learning and backpropagation circuits create an overhead for additional memory and registers to store intermediate outputs for gradient calculation, control circuits and flexible peripheral circuits for a crossbar array. Moreover, neural network training still requires higher bit precision than inference, which also leads to higher resolution ADCs. The development of efficient and flexible hardware support for on-chip training is an open challenge.

In addition, the QNN IMC hardware should support state-of-the-art software-based training techniques. For example, normalization fusion of neural network layers \cite{jung2019restructuring}, where batch normalization layers are merged into the previous fully-connected or convolutional layers. Implementation of such a method on IMC hardware might bring accuracy degradation if the trained neural network is directly quantized and deployed on IMC accelerators. Overall, it is important to transfer the software-based methods for QNN to IMC hardware.

\subsection{IMC Hardware Support for Mixed-precision Designs and Reconfigurable Precision}

Recently, mixed-precision quantization has become an essential part of IMC hardware design, as it is more effective than fixed-precision quantization \cite{huang2021mixed}. The IMC hardware support for mixed-precision designs is an open challenge.
This implies that IMC macros should support a different number of bits for weights and activations, and have the ability to be reconfigured for different precision schemes, which also leads to hardware overhead related to computational blocks and control circuits.
 Moreover, the hardware under-utilization problem should be considered in such architectures, and reconfiguration schemes for effective hardware resource utilization in IMC macros are required for mixed-signal designs.

\subsection{Moving Towards General-Purpose Architectures by Improving IMC Design Reconfigurability}

Most of the hardware for IMC architectures is designed for specific applications and specific QNN types. However, the overall trend in IMC designs is to move towards general-purpose chips \cite{zhang2020neuro}.
To make a step towards general-purpose functionality, apart from the support of reconfigurable precision, IMC designs should support different workloads, different directions of data movement, quantization schemes, data types, and both spatial and temporal domain computations.
The distribution of the workloads in different types of neural networks varies, e.g. ResNet workloads and data distribution for pattern recognition are different from LSTM workloads for language processing. Moreover, there is also a variation of the workloads within a neural network for different layers, e.g. convolution layers and fully-connected layers process the data differently and IMC hardware should support reconfigurability to be programmed accordingly. 
The issue of different workloads leads to unbalanced execution time and resource utilization in different layers \cite{zhang2020neuro}.

Reconfigurability requires the hardware support of flexible interconnects.
The support of different types of quantization schemes is important, as different applications rely on different data distributions. Moreover, the support of corresponding arithmetic operations and related IMC hardware macros is essential. 
At the same time, it is important to identify the trade-off between the support of different operations and quantization schemes and related hardware overhead.
Overall, to satisfy different tasks, network structures, quantization methods, precision, and latency requirements, IMC QNN architectures should be reconfigurable. 

Reconfigurability is also useful for balancing energy consumption and performance accuracy. 
There is a trade-off between the resources required for the task execution and the performance accuracy in
low-power on-edge designs. 
In general, low-precision architectures usually lead to lower energy consumption but lower accuracy. Flexible precision and reconfigurability of IMC architectures can be used for scenarios when the power source is not available for on-edge devices, so the application can run under constrained power resources at the cost of reduced performance accuracy.

\subsection{Automating the Quantization Policies Search in IMC Architectures}

Quantization policies have several parameters to optimize to deploy to IMC hardware, including bit widths, scaling factors, quantization thresholds, and clipping ranges. This problem becomes even more complex for mixed-precision quantization, where different layers can have different quantization policies. Moreover, when deploying on IMC hardware, the performance accuracy can also be affected by IMC hardware non-idealities \cite{krestinskaya2020automating}. 
Therefore, it is important to include hardware-related evaluation when optimizing quantization policies, as it can affect hardware performance accuracy. 
It is difficult to find the optimum policies manually, especially when the size of the network increase. To automate the search for optimum quantization policies, it can be formulated as Hardware-Aware Neural Architecture Search (HW-NAS) problem. Such an approach is presented in \cite{huang2021mixed}, where optimum bit widths for weights and precision for ADC for IMC architectures are searched using a reinforcement learning approach. CMQ framework \cite{peng2022cmq} also searches for optimum quantization threshold and bit width using a differentiable search approach. Gibbon framework \cite{sun2022gibbon} searches for optimum bit width and ADC precision along with neural network architecture and crossbar-related hardware parameters using an evolutionary algorithm-based approach. The further development of such frameworks is useful, as there are no unified frameworks considering the joint optimization of neural network architecture parameters, hardware parameters, and quantization policies. Moreover, the existing frameworks are designed for specific architectures and specific problems and require modifications for general use. Therefore, the development of a unified joint optimization framework for HW-NAS including quantization policies optimization is an open challenge.

\subsection{Software-Hardware Co-design}

Software-Hardware Co-design connects a software-based design of a neural network to hardware its implementation and implies the optimization of all levels of design from the devices and circuit macros to architectures and algorithms \cite{zhang2020neuro}.
Currently available frameworks for IMC architectures, e.g. PUMAsim \cite{ankit2019puma} or NeuroSim \cite{peng2019dnn+}, focus on a specific IMC design, and do not support different quantization techniques, especially mixed-precision quantization, and have limited support of various IMC devices.
There is still a lack of a universal EDA toolchain for large-scale implementations supporting various IMC hardware designs \cite{zhang2020neuro}, including different types of neural networks, IMC architectures with various IMC devices and related non-idealities, various macros, and different designs of interconnects for the architecture blocks. 
Software-Hardware Co-design frameworks should include more QNN features, support different quantization methods, and include more mapping techniques.
Overall, the efficient automated compiler for mapping QNN to an IMC-based hardware implementation supporting a wide range of IMC designs is still an open challenge.

\subsection{Integration of IMC Architectures to Traditional Hardware Designs and Combining Different Types of Hardware}

While IMC architectures are efficient for certain operations, e.g. MVMs, they cannot fully replace the other types of hardware. 
Therefore, it is important to integrate IMC architectures with traditional hardware designs and different types of hardware. In particular, this approach will also benefit QNN training on a chip, where floating point operations are still necessary and require different hardware or additional DSP blocks. The efficient integration of various types of IMC hardware together is also necessary. 
There has been a successful attempt to combine an RRAM-based IMC accelerator with conventional SRAM-based memories and embedded processor \cite{chang202240nm}. 
The efficient integration of non-volatile memories with traditional memories, e.g. SRAM, is important to overcome external memory requirements and latency issues. 
Large-scale integration of relatively novel IMC architectures with traditional hardware designs can bring a lot of benefits in terms of architecture efficiency and is the next step in IMC hardware development.

\section{Conclusion}

This paper reviews state-of-the-art designs of IMC-based QNN hardware implementations and compares the QNN designs with different IMC devices. To improve the efficiency of IMC-based QNNs, different levels of the design from IMC devices and architectures to QNN algorithms should be improved simultaneously. The main challenges and future directions in IMC-based QNN hardware research include further improvement of the IMC-based inference engines for QNNs, efficient on-chip training with quantized gradients and weight updates, hardware support of mixed-precision quantization, reconfigurability of the designs, automation of the optimum quantization policies search along with optimized hardware parameters, software-hardware co-design, and integration of IMC architectures to traditional hardware designs.

% or


%\appendices

%\section{}

%\section*{Acknowledgment}


\ifCLASSOPTIONcaptionsoff
  \newpage
\fi



%\bibliographystyle{IEEEtran}

\vspace{-0.15cm}
\setstretch{0.8}

\bibliographystyle{IEEEtran}
\bibliography{ref}



%\newpage



%\vfill


\end{document}